\documentclass[%
 reprint, 
superscriptaddress,
 amsmath,amssymb,
 aps, physrev,
floatfix,
]{revtex4-2}

\usepackage{graphicx}
\usepackage{dcolumn}
\usepackage{bm}
\usepackage{multirow}
\usepackage{makecell}
\usepackage{array}
\usepackage{float}
\usepackage{placeins}
\usepackage[separate-uncertainty=true,multi-part-units=single]{siunitx}
\usepackage[hyperindex,breaklinks,hidelinks]{hyperref}

\begin{document}


\title{Drive beam depletion with multi-Joule energy transfer in a plasma wakefield accelerator}

\author{R. Ariniello}
\email{robert.ariniello@colorado.edu}
\affiliation{SLAC National Accelerator Laboratory, Menlo Park, California 94025, USA}
\affiliation{Center for Integrated Plasma Studies, Department of Physics, University of Colorado Boulder, Boulder, Colorado 80309, USA}
\author{V. Lee}
\affiliation{Center for Integrated Plasma Studies, Department of Physics, University of Colorado Boulder, Boulder, Colorado 80309, USA}
\author{D. Storey}
\affiliation{SLAC National Accelerator Laboratory, Menlo Park, California 94025, USA}
\author{C. Emma}
\affiliation{SLAC National Accelerator Laboratory, Menlo Park, California 94025, USA}
\author{S. Gessner}
\affiliation{SLAC National Accelerator Laboratory, Menlo Park, California 94025, USA}
\author{M. J. Hogan}
\affiliation{SLAC National Accelerator Laboratory, Menlo Park, California 94025, USA}
\author{A. Knetsch}
\affiliation{SLAC National Accelerator Laboratory, Menlo Park, California 94025, USA}
\author{M. D. Litos}
\affiliation{Center for Integrated Plasma Studies, Department of Physics, University of Colorado Boulder, Boulder, Colorado 80309, USA}
\author{N. Majernik}
\affiliation{SLAC National Accelerator Laboratory, Menlo Park, California 94025, USA}
\author{B. O'Shea}
\affiliation{SLAC National Accelerator Laboratory, Menlo Park, California 94025, USA}

\date{\today}

\begin{abstract}
A collider based on beam-driven plasma wakefield acceleration will require the drive beam to transfer 10-100s of Joules to the plasma in each stage, with a drive-to-wake energy transfer efficiency exceeding $70\%$. Using an all-optical plasma source, we demonstrate significant progress towards these parameters, transferring at least \qty{5.6}{J} from a \qty{1.52}{nC}, \qty{10}{GeV} electron beam to a \qty{4.5e16}{cm^{-3}} hydrogen plasma while achieving at least $37\pm 3\%$ drive-to-wake energy transfer efficiency. We observe deceleration of some particles to less than \qty{0.93}{GeV} with up to 90\% of the charge participating in the interaction. 
\end{abstract}

\maketitle

\section{\label{sec:intro}Introduction}

Plasma wakefield acceleration (PWFA) can generate accelerating gradients three orders of magnitude larger than conventional accelerators \cite{hogan2005multi,blumenfeld2007energy}, opening the potential for TeV scale lepton colliders for high-energy particle physics \cite{adli2013beam,foster2023hybrid,Asai:2905898}. In beam-driven PWFA, an ultra-relativistic drive beam excites a density wave in a plasma, called a wake, producing strong transverse and longitudinal electromagnetic wakefields that accelerate and focus a trailing witness bunch. These strong wakefields result in the witness beam gaining---and the drive beam losing--- significant energy over a short distance. The use of wakefield acceleration promises to reduce the construction cost of a collider by reducing the facility footprint, but to minimize operating costs, the accelerator must also have high wall-plug efficiency. Achieving this high efficiency is only possible by maximizing the efficiency of each step in the energy transfer chain: wall-plug to rf, rf to drive beam, drive beam to plasma wake, and plasma wake to witness beam. Much progress has been made on producing electron beams efficiently with conventional linacs through the development of the CLIC proposal \cite{aicheler2014multi}. Klystrons with wall-plug to rf efficiencies up to 65\% are commercially available, and efficiencies up to 90\% have been proposed \cite{baikov2015toward}, while 95\% rf to drive beam efficiency has been demonstrated at CTF3 leading to wall-plug to drive beam efficiencies as high as 55\% \cite{aicheler2014multi,Urschutz:2006wgn}. 

On the plasma wakefield side, a drive-to-wake efficiency of 57\% \cite{pena2024energy} and a wake-to-witness efficiency of 42\% have been demonstrated \cite{litos2014high, lindstrom2021energy}. These efficiencies are promising, but still less than the values specified in existing PWFA collider design studies; HALHF assumes 72\%  for drive-to-wake \cite{foster2023hybrid}, while Ref. \cite{adli2013beam} assumes 76\%, both based off idealized simulations. The demonstrated efficiency is limited by the re-acceleration of electrons that have lost sufficient energy such that they are no longer ultra-relativisitic and slip back into the accelerating phase of the wake. Mitigating this re-acceleration requires going to larger drive beam energies. In addition, higher drive beam energy and charge are necessary for a collider to minimize the number of stages required. As a comparison, the 57\% of Ref. \cite{pena2024energy} was achieved using a \qty{636}{pC}, \qty{0.5}{GeV} drive beam while HALHF and Ref. \cite{adli2013beam} assume, respectively, \qty{4.27}{nC} at \qty{31.25}{GeV} and \qty{3.2}{nC} at \qty{25}{GeV}. Scaling to these collider parameters is a challenge because the total energy (energy per unit length) deposited in the plasma are up to two orders of magnitude higher: \qty{96}{J} (\qty{19.2}{J/m}) for HALHF and \qty{60.8}{J} (\qty{18.4}{J/m}) for Ref. \cite{adli2013beam}, compared to the \qty{0.18}{J} (\qty{0.92}{J/m}) of Ref. \cite{pena2024energy}. Increasing the energy deposited can lead to difficulties with heat dissipation in the plasma source. For example, the alkali heat pipe oven at the Facility for Advanced Experimental Tests (FACET-II) has suffered overheating and lithium loss when trying to increase the drive-to-wake energy transfer with a \qty{10}{GeV} drive beam.

In this paper, we avoid these problems by using a \qty{10}{GeV} electron beam in a plasma formed by a novel pair of diffractive optics that laser ionize a \SI{4.5e16}{cm^{-3}} hydrogen gas to form a \qty{85}{cm}-long, \qty{200}{\micro\meter}-diameter plasma source \cite{ariniello2025tandemlens}. In this source, we demonstrate over an order of magnitude improvement in energy deposited, reaching \qty{5.6}{J} (\qty{6.6}{J/m}) at \qty{10}{Hz} repetition rate. We observed that over 90\% of the charge participated in driving the wake, allowing us to achieve at least $37\pm 3\%$ drive-to-wake energy transfer efficiency. On some shots, energy depletion below the spectrometer's lower limit of \qty{0.93}{GeV} was observed for an appreciable fraction of the particles in the beam. Our results are a significant step forward towards the parameters required for a linear collider.

\section{\label{sec:exp}Experimental Setup}

\begin{figure*}[bt]
  \centering
  \includegraphics[width=7in]{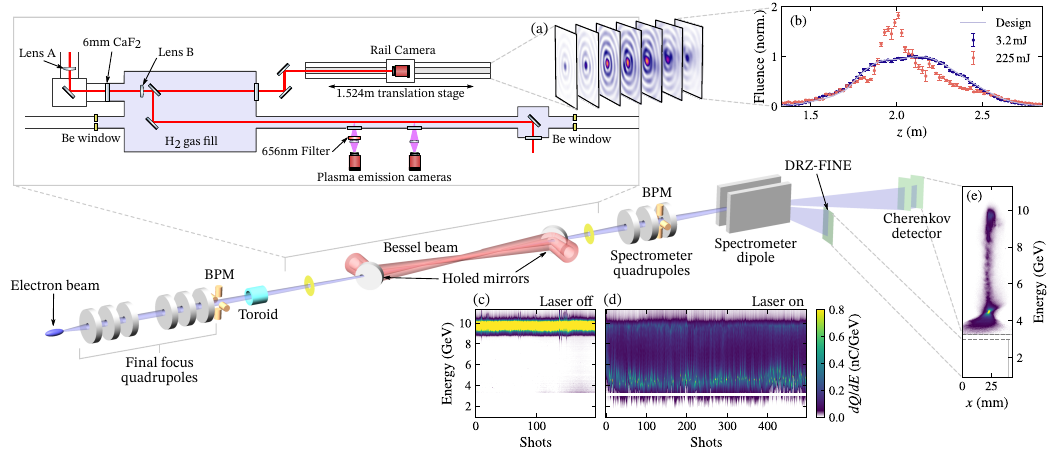}
  \caption{Experimental setup. The laser pulse reflected off a holed mirror to co-propagate with the electron beam where it ionized the $\mathrm{H_2}$ gas filling the beamline. The gas was contained between two holed Be windows by a differential pumping system. The FACET-II electron beam was focused into the plasma with a pair of quadrupole triplets. After the plasma, the beam was imaged by three quadrupoles into a dipole spectrometer. (a) The laser focal region was characterized by scanning a camera (Rail camera) through the leakage light from a turning mirror. (b) The on-axis fluence was distorted at high power by nonlinear phase picked up in the $\mathrm{CaF_2}$ that isolates the laser compressor from the beamline. (c) Electron beam spectra with the laser off, showing minimal interaction between the beam and the $\mathrm{H_2}$ gas. (d) With the laser on, significant deceleration due to the laser ionized plasma was visible. Shots were recorded sequentially, with pauses for saving data every 100 shots. (e) A representative shot [shot 15 in (d)] is shown, measured using two detectors: a scintillator screen for particles from \qty{0.93}{GeV} to \qty{3}{GeV} and a Cherenkov based detector for particles from \qty{3.25}{GeV} to \qty{11.30}{GeV}.  \label{fig:01-ExperimentalSetup}}
\end{figure*}

The PWFA experiment took place at the FACET-II facility at SLAC National Accelerator Laboratory \cite{yakimenko2019facet}. Electron bunches were generated by a photocathode injector then alternately accelerated and compressed in three stages to deliver \qty{10}{GeV}, \qty{1.52}{nC}, $<\qty{20}{\micro\meter}$ long bunches to the experimental area. A final focus, composed of two quadrupole triplets, focused the beams to the entrance of the plasma with a \qty{50}{cm} beta function. The beams had a spot size at the focus of \qty{\sim30}{\micro\meter}, measured by wirescanner. A laser heater in the FACET-II injector was used to suppress micro-bunching in the beam that can lead to beam ionization of hydrogen \cite{huang2004suppression, zhang2024generation}.

The plasma source was formed by filling the interaction point (IP) region with \qty{1.35}{torr} of hydrogen gas and singly ionizing the $\mathrm{H_2}$ molecules with a \qty{800}{nm}, \qty{225}{mJ}, \qty{45}{fs} FWHM laser pulse (Fig.~\ref{fig:01-ExperimentalSetup}) \cite{green2014laser}. The gas was contained in a \qty{4}{m} long region by the inner two apertures of a differential pumping system, which separated the linac vacuum from the IP without any beam-intercepting windows \cite{storey2024wakefield}. A continuous gas inflow replaces the outflow through the apertures, maintaining the pressure to within 1\% of the set-point. The laser was focused by a pair of diffractive optics (Lens A and B in Fig.~\ref{fig:01-ExperimentalSetup}) designed to produce an \qty{1}{m} long Bessel focus with a width of \qty{120}{\micro\meter} (the diameter of the first Bessel function 0). The laser system was isolated from the hydrogen gas by a \qty{6}{mm} thick $\mathrm{CaF_2}$ window. The laser fluence through the focal region was measured by scanning a camera along the length of the focus. The results are shown in Fig.~\ref{fig:01-ExperimentalSetup}(a-b), where $z$ is measured from Lens B and 10 images were taken at each $z$ step, with the average and standard deviation shown. The diffractive optics were designed to produce a controlled on-axis plasma density, for details see Ref.~\cite{ariniello2025tandemlens}. A camera---imaging the beam axis through a window on the side of the beamline---confirmed plasma formation by detecting light emission from the H-alpha line through a \qty{656.3}{nm}, \qty{10}{nm} FWHM line filter. 

The primary beam diagnostic after the plasma stage was an imaging spectrometer, comprised of three quadrupoles for imaging and a dipole to disperse the beam vertically in energy. Two detectors were used to view the dispersed beam: a Cherenkov radiation based detector located \qty{10.28}{m} downstream of the dipole measured electrons dispersed between 4.6-\qty{16.1}{mrad}, and a DRZ™-FINE scintillating screen located \qty{4.65}{m} downstream of the dipole measured electrons dispersed from 17.7-\qty{57.3}{mrad} [Fig.~\ref{fig:01-ExperimentalSetup}(e)]. The dipole strength was varied to change the energy range visible on the two screens. The incoming charge was measured on a shot-to-shot basis using a BPM before the plasma, which was calibrated against a nearby toroid. 

Detailed descriptions of the spectrometer and detectors are provided in Refs. \cite{storey2024wakefield, adli2015cherenkov}. Energy calibrations for both detectors were acquired by varying the dipole strength while measuring the position of the \qty{10}{GeV} beam on the screen. The charge response of the Cherenkov detector as a function of vertical position was determined by recording the counts on the detector and the beam charge on a toroid while scanning the beam vertically with the dipole. The \qty{1.52}{nC} beam saturates the camera viewing the scintillator screen; an approximate charge calibration was instead found using the portion of the beam decelerated by the PWFA. The spectrometer was set to image at the energy in the gap between the two detectors---to avoid charge loss at those energies---and the charge calibration was determined by requiring that the spectrum be continuous across the gap. The uncertainties in the charge and energy calibration are included in the errors presented for charge and energy measurements. 

Temporal overlap between the laser and the electron beam was achieved by electronically delaying the laser until the cessation of beam-plasma interaction on the spectrometer. The laser was then set to arrive \qty{5}{ps} before the electron beam, much less than the \qty{\sim1}{ns} timescale of the expansion of the plasma \cite{lee2024temporal}. The plasma source was aligned to the electron beam by looking at the plasma afterglow light on two cameras at different longitudinal positions along the beamline, the procedure is detailed in Ref.~\cite{Lee2025alignment}. After alignment, the PWFA interaction was readily apparent on the spectrometer. The interaction was optimized by scanning the rf phase of the accelerating section before the second bunch compressor. This changed the bunch length and current profile after compression. Further optimization was performed by scanning the beam waist longitudinally using the final focus magnets.

\section{Estimate of Plasma Length}

The $\mathrm{CaF_2}$ window in the laser path imparts some non-linear phase on the laser pulse. At \qty{3.2}{mJ} of laser energy, the nonlinear effects are negligible and the on-axis fluence matches the design expectations. At \qty{225}{mJ}, self-focusing in the window results in a distorted profile as shown in Fig.~\ref{fig:01-ExperimentalSetup}(b). Even with the distortion, the laser produced a sufficiently large plasma for the PWFA experiments, but the length of the plasma differed from the design length. No in-situ diagnostics were available to measure the plasma density along the beamline. Instead, we used the electron bunches delivered by the FACET linac as a probe of the plasma.

\begin{figure}[bt]
  \centering
  \includegraphics[width=3.37in]{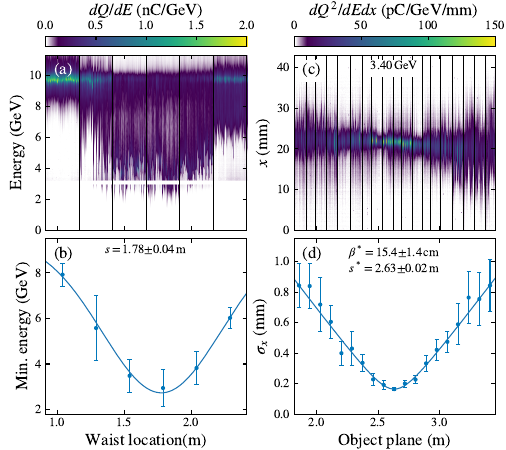}
  \caption{Estimate of the plasma length. (a) Energy spectra recorded from scanning the location of the incoming beam waist at the plasma entrance, black lines separate the scan steps. (b) The energy loss is maximized when the waist is positioned at the start of the plasma, as this position minimizes emittance growth. (c) Transverse projection of a \qty{20}{MeV} energy slice centered at \qty{3.4}{GeV} as the spectrometer imaging plane is scanned. (d) Fitting Eq.~(\ref{eq:vacProp}) gives both the $\beta$ function and location of the waist at the exit of the plasma. \label{fig:PlasmaLength}}
\end{figure}

The optimum waist location approximately corresponds to the start of the plasma. This is because the interaction is maximized when the waist is co-located with the virtual-waist of the plasma's matched beta function \cite{zhao2020emittance}, which corresponds to approximately the center of the plasma entrance ramp. Figure~\ref{fig:PlasmaLength}(a) shows the energy spectrum of the beam through the waist scan. The longitudinal waist location with maximum interaction was found by fitting a Gaussian, Fig.~\ref{fig:PlasmaLength}(b), to the maximum energy loss. The start of the plasma was measured to be at $z\approx\qty{1.78}{m}$ downstream of Lens A, only slightly different from the start expected at $z=\qty{1.67}{m}$ in the absence of laser distortion.

The location of the end of the plasma was found by measuring the virtual-waist of the beam exiting the plasma. The spectrometer magnets were scanned to point-to-point image \qty{3.5}{GeV} electrons across a range of object planes. The scan range captured the virtual-waist for each energy slice between \qty{3.3}{GeV} and \qty{3.6}{GeV}. This energy range of the recorded spectra were sliced into \qty{20}{MeV} bins and the beam width in each bin was found by fitting a Gaussian. Dividing by the spectrometer magnification for each energy slice and quad setting gave the beam width at each object plane. Shots with poor fits ($R^2<0.9$) were rejected. The beta function at the virtual-waist $\beta^*$ and waist location $s^*$ were found by fitting the vacuum propagation equation
\begin{equation} \label{eq:vacProp}
  \sigma_x=\sqrt{\epsilon[\beta^* + (s-s^*)^2/\beta^*]}
\end{equation}
to the measured beam sizes, as shown in Fig.~\ref{fig:PlasmaLength}(b). For particles at \qty{3.4}{GeV}, the exit waist was determined to be located at $z=\qty{2.63\pm0.02}{m}$, with a beta function of \qty{15.4\pm1.4}{cm} and an emittance of \qty{1.16\pm0.08}{mm-rad}. The location of the virtual waist varied across the energy slices from \qty{2.60}{m} to \qty{2.65}{m}, some variation is expected from transverse theory \cite{ariniello2019transverse,ariniello2022chromatic}, but the exact amount depends on the details of the bunch's longitudinal structure, which is not well known here.

\begin{figure*}[bt]
  \centering
  \includegraphics[width=7in]{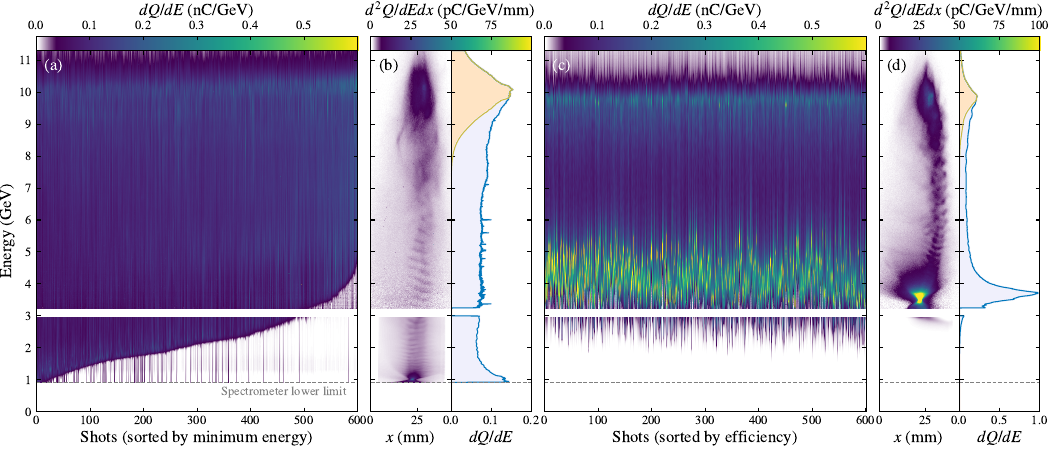}
  \caption{Energy depletion of the electron beam (a and b) and high drive-to-wake energy transfer efficiency (c and d). (a) Spectrum of all shots taken with the spectrometer set to image $\leq\qty{1.5}{GeV}$, sorted by minimum energy. (b) An example spectrometer image and spectrum (shot 0) showing depletion of significant charge below the spectrometer's lower limit of \qty{0.93}{GeV}. The orange shading depicts the amount of charge that remains at \qty{10}{GeV} and does not participate in driving the wake, here $236\pm\qty{3}{pC}$. (c) The 600 shots with the greatest drive-to-wake energy transfer efficiency, taken with the spectrometer set to image \qty{3.5}{GeV}. (d) Spectrum of a shot with high efficiency (shot 2, $35\pm 3\%$ drive-to-wake efficiency) showing a significant fraction of the charge located in a single peak around \qty{3.75}{GeV}. \label{fig:EnergyDepletion}}
\end{figure*}

Using the average exit position of all energy slices, the length of the plasma is approximately \qty{0.85}{m} resulting in a gradient of \qty{10.7}{GeV/m} for particles at the spectrometer lower limit.

\section{Drive Depletion and Energy Transfer}

We looked for the depletion of drive beam electrons by setting the dipole strength as low as possible while keeping the \qty{10}{GeV} beam visible in the Cherenkov detector's field of view. With this dipole setting, the scintillator screen detects electrons between \qty{0.93}{GeV} and \qty{3.00}{GeV} while the Cherenkov detector detects electrons between \qty{3.25}{GeV} and \qty{11.30}{GeV}. The spectrometer quadrapoles were set to image the exit of the plasma to the scintillator screen. The energy imaged by the spectrometer was scanned from \qty{3}{GeV} to \qty{1}{GeV} to find the minimum energy visible. With the imaging energy set to $\leq\qty{1.5}{GeV}$, charge was observed down to the minimum energy measurable on the screen. Figure~\ref{fig:EnergyDepletion}(a) shows the spectrum of all shots recorded with imaging energy set to $\leq\qty{1.5}{GeV}$. The individual spectra have been sorted by the minimum energy with spectral charge density $\frac{dQ}{dE}>\qty{45}{pC/GeV}$. Approximately 20 shots were recorded with charge down to the minimum energy measurable. The charge depleted below the spectrometer limit was considerable on several shots, such as that shown in Fig.~\ref{fig:EnergyDepletion}(b). 

A lower bound on the amount of charge at low energy can be estimated by considering the source of charge loss in the beamline. With the spectrometer set to image $\leq\qty{1.5}{GeV}$, 50-76\% of the charge entering the plasma was visible on the spectrometer diagnostics across all the shots. The missing charge was lost in three places: (i) charge that transversely missed the edges of the detector's field of view, (ii) charge that was lost in transport between the plasma and detectors (due to the large divergence of low energy particles from the plasma \cite{pena2024energy} or large $M_{12}$ for particles far from the imaging energy), and (iii) charge that fell in the energy gap between the spectrometer detectors. We can estimate a lower bound for (ii) by comparing the incoming charge to the charge measured on a BPM after the second spectrometer quadrupole. For shots with minimum electron energy $>\qty{1.5}{GeV}$, only $10.8\pm\qty{9.4}{pC}$ of charge on average was missing after the second quadrupole, where as $122\pm\qty{1.5}{pC}$ was missing for the shot shown in Fig.~\ref{fig:EnergyDepletion}(b) implying that there was at least \qty{100}{pC} of additional charge with energy $<\qty{1.5}{GeV}$ that never reached the spectrometer. In addition to the significant amount of charge approaching depletion, the non-participating charge at the front of the wake was minimal. For shots close to energy depletion (minimum energy less than \qty{2}{GeV}), on average only $14\pm2\%$ of the incoming charge did not participate in driving the wake.

As the accelerator drifted away from the optimal set point, the minimum energy observed increased, allowing more charge to be captured on the spectrometer screens. To observe this charge, the spectrometer quadrupoles were set to image \qty{3.5}{GeV} particles onto the Cherenkov detector, see Fig.~\ref{fig:EnergyDepletion}(c), allowing the spectrometer to capture up to $94\pm 2\%$ of the incoming charge on shots where the minimum energy was close to the imaging energy. A lower bound on the drive-to-wake energy transfer efficiency was found by assuming the missing charge had the same per-particle-energy as the incoming beam. The best shot had an efficiency of $35\pm 3\%$. Some charge is lost in the gap between the spectrometer screens; using linear interpolation to reconstruct the spectrum in this region increases the minimum bound on efficiency to $37\pm 3\%$. The incoming beam contained \qty{15.2}{J} of energy, corresponds to at least \qty{5.6}{J} of energy deposited in the plasma, or \qty{6.6}{J/m} over the \qty{0.85}{m} plasma length.

The spectrum of the decelerated electrons, which is shown for a representative shot in Fig.~\ref{fig:EnergyDepletion}(d), has excellent quality, with most of the decelerated charge located in a narrow peak near the minimum energy. Such a spectrum implies that $>60\%$ energy transfer efficiency can be achieved, without significant re-acceleration, by lengthening the plasma to nearly fully deplete the electrons in the low energy peak. 


\section{\label{sec:conc}Conclusion}

We have experimentally demonstrated multi-Joule energy transfer and pump depletion in a PWFA by using a laser ionized plasma source that is insensitive to overheating. Our results here are within an order of magnitude of the drive-to-wake energy transfer required for a collider, and within a factor of three of the energy deposited per unit length. Scaling up by at least a factor of two should be possible by increasing the plasma length to fully deplete the beam and increasing the drive beam charge to \qty{2.1}{nC}, which is the maximum charge FACET-II should eventually be able to deliver. With these changes, efficiency should exceed $60\%$ and energy deposited would more than double to \qty{12.6}{J}. The full depletion results here suggest that even with the existing plasma length, it is likely possible to achieve efficiencies of $>60\%$ leading energy deposition of \qty{15}{J/m}, comparable to that proposed for a collider and the near the maximum possible from the FACET-II facility. This will require improvements in beam stability to allow for scanning the imaging energy of the spectrometer to capture all of the decelerated charge. 

In the process of carrying out the experiment, PWFA interaction was sustained nearly continuously for nearly \qty{8}{hours}, demonstrating that the plasma source does not suffer any long term heat build-up.


\begin{acknowledgments}

This work was supported by the U.S. Department of Energy under DOE Contract No. DE-AC02-76SF00515 and by the U.S. Department of Energy, Office of Science, Office of High Energy Physics under Award No. DE-SC0017906.

\end{acknowledgments}

\bibliography{bibliography}

\end{document}